\documentclass[a4paper]{article}

\usepackage{INTERSPEECH2021}

\usepackage{color}
\usepackage{dsfont}
\usepackage{amsmath,amssymb}
\definecolor{grey}{rgb}{0.5, 0.5, 0.5}
\newcommand\spv[1]{\textcolor{grey}{#1}}

\newcommand{\omitme}[1]{}

\title{Multimodal Self-Supervised Learning of General Audio Representations}
\name{Luyu Wang, Pauline Luc, Adri\`a Recasens, Jean-Baptiste Alayrac, A\"aron van den Oord}
\address{DeepMind}
\email{\{luyuwang, paulineluc, arecasens, jalayrac, avdnoord\}@google.com}

\begin{document}

\maketitle
\begin{abstract}
  We present a multimodal framework to learn general audio representations from videos.
  Existing contrastive audio representation learning methods mainly focus on using the audio modality alone during training.
  In this work, we show that additional information contained in video can be utilized to greatly improve the learned features.
  First,  we demonstrate that our contrastive framework does not require high resolution images to learn good audio features. 
  This allows us to scale up the training batch size, while keeping the computational load incurred by the additional video modality to a reasonable level.
  Second, we use augmentations that mix together different samples. We show that this is effective to make the proxy task harder, which leads to substantial performance improvements when increasing the batch size. 
  As a result, our audio model achieves a state-of-the-art of 42.4 mAP on the AudioSet classification downstream task, closing the gap between supervised and self-supervised methods trained on the same dataset.
  Moreover, we show that our method is advantageous on a broad range of non-semantic audio tasks, including speaker identification, keyword spotting, language identification, and music instrument classification.
\end{abstract}

\section{Introduction}
Self-supervised learning has recently emerged as an alternative to supervised learning doing away with the requirement for manually annotated labels \cite{oord2018representation, bachman2019learning, chen2020simple, he2020momentum}. It can leverage large amounts of unlabelled data and produce competitive performance on vision, language, and speech tasks \cite{henaff2020data, miech2020end, alayrac2020self, radford2021learning, devlin2019bert, schneider2019wav2vec, kawakami2020learning, baevski2020wav2vec}. The constrastive learning framework has attracted a great amount of attention due to its strong performance on image tasks \cite{bachman2019learning, chen2020simple, he2020momentum}. It relies on a Siamese architecture \cite{bromley1993signature}, in which two views are created with predefined augmentations. Semantically similar samples (positives) are brought closer in the feature space and dissimilar ones (negatives) are pushed apart using a contrastive loss.
Several approaches~\cite{miech2020end, alayrac2020self, radford2021learning} use similar contrastive objectives, adapted to learn video representations from the supervision brought by the text and/or audio modalities.
Although the progress has been rapid, these works are vision-centric, in that design decisions are made based on the performance of the video representations. In \cite{saeed2020contrastive, fonseca2020unsupervised, wang2021multi} however, the contrastive framework has been shown to be very effective on learning audio representations. Positive pairs are constructed from augmented audio segments from the same clip while negatives come from different ones. In particular, \cite{wang2021multi} proposes a multi-format approach: in contrast to the Siamese setup employed by contrastive approaches, for each sample, a spectrogram representation and a waveform representation are extracted from the audio modality alone, which are then independently augmented and processed by two different networks: a \emph{spectrogram network} and a \emph{waveform network}. In this setup, each network builds a representation for its respective input format via the supervision provided by the other one.
\omitme{
shows that training two different networks, in which one takes the raw audio as the input and the other uses the spectrograms, can substantially improve performance of the learned features. Even though it considers two formats, it only uses the audio modality. It remains unclear what is the role of the video modality on learning audio representations under the contrastive learning framework.}

Meanwhile, other forms of the self-supervised objective have also been considered for learning sound representations \cite{arandjelovic2017look, jansen2018unsupervised, jansen2019coincidence, tagliasacchi2020pre, wang2020contrastive, shor2020towards}.
In \cite{jansen2018unsupervised, shor2020towards}, a triplet loss is employed to minimize the distance between the anchor and positives. Hard negative mining is usually required to successfully train with this objective.
\omitme{temporally close positives augmented with noise injection and temporal and frequency shifts.}
It has later been extended to multimodal setting with an additional clustering-based loss, and an improvement on the AudioSet tagging task has been reported \cite{jansen2019coincidence}.
While these methods usually take audio spectrograms as the input, \cite{wang2020contrastive} shows that Contrastive Predictive Coding (CPC) is effective to learn audio representations from raw waveforms.

\begin{figure}[t]
  \centering
  \includegraphics[width=0.95\linewidth]{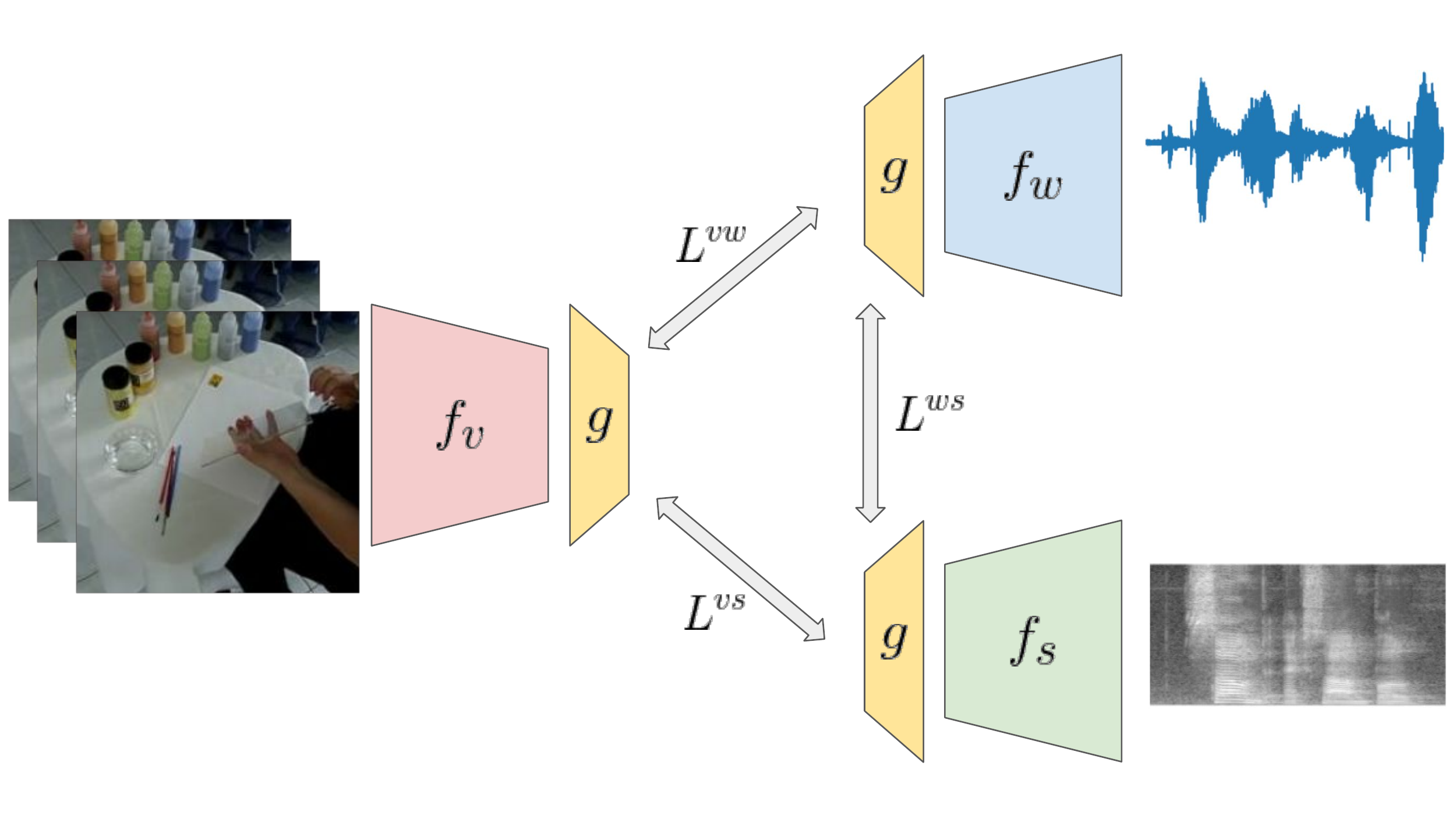}
  \caption{Illustration of our contrastive learning framework that takes videos, waveforms, and spectrograms as inputs. Details can be found in Section~\ref{sec:method}.}
  \label{fig:framework}
\end{figure}

In this work, we push the limits of constrastive learning of audio representations with the aid of the video modality, building on the multi-format approach~\cite{wang2021multi}. Unlike in vision \cite{miech2020end, alayrac2020self}, we find that the resolution of the video is not crucial to learn strong audio representations. This observation largely decreases the computational cost incurred by the use of the video modality.
We show that an augmentation procedure which mixes together different samples is very effective to make the discriminative pretraining task harder, which translates into significant performance improvements as we increase the batch size.
On the AudioSet benchmark \cite{hershey2017cnn}, our spectrogram network achieves a new state-of-the-art mAP of 42.4, approaching the performance of the supervised method trained on the same dataset (mAP 43.1) \cite{kong2020panns}.
Meanwhile, our waveform network, with an mAP of 40.5, even outperforms the supervised state-of-the-art at 36.5 \cite{kong2020panns}. Furthermore, we show that our method outperforms prior work \cite{saeed2020contrastive} on a broad class of downstream tasks, including speaker identification, language identification, music instrument classification, hot word spotting, and several others, showing the generality of the learned representations. Our results suggest videos can provide strong supervisory signal in the absence of labels.

\section{Multimodal Contrastive Audio Learning Framework}
\label{sec:method}
Our contrastive learning framework is depicted in Figure~\ref{fig:framework}.
For self-supervision, we take log-mel spectrograms (S), raw waveforms (W), and RGB frame sequences (V) as inputs.
Therefore, there are three encoder networks involved, namely, the \emph{spectrogram network} $f_s$, \emph{waveform network} $f_w$, and \emph{video network} $f_v$.

At training time, we randomly sample two crops from the raw audio of a training video sample, and one set of image frames synchronized to the first audio crop. Then we extract spectrograms from the first audio crop, and keep the second one as a raw waveform. Before feeding them to the encoders, we first augment each input modality independently. For videos, we use random spatial cropping, scale, and color jittering. For spectrograms, we apply a truncated shift in frequency by an integer number sampled from $[-F, F]$, where $F$ is the maximum shift size. Missing values after the shift are set to 0. For waveforms, we do not use format-specific augmentations.

We further use example mixing \cite{zhang2017mixup} to augment each type of input.
Given two inputs $\mathbf{x_1}$ and $\mathbf{x_2}$ from the batch, the mixed-up version of $\mathbf{x_1}$ is
\begin{equation}\label{eq2}
  \mathbf{\hat{x}_1} = \alpha \mathbf{x}_1 + (1 - \alpha) \mathbf{x}_2,
\end{equation}
where $\alpha$ is the mixing ratio controlling the strength of the distractor $\mathbf{x_2}$.
Audio mixing has been found to be very effective under both the supervised \cite{kong2020panns} and unsupervised setting \cite{jansen2018unsupervised, wang2021multi}. In this work, we also investigate the use of this technique for videos.

We use pairwise contrastive losses to construct our objective.
A positive pair $\left(\mathbf{x}^a_i, \mathbf{x}^b_i\right)$ is created from cropping a given training sample's modalities $a$ and $b$, respectively.
On the contrary, negative pairs are constructed from different samples.
Next, the networks encode the pairs augmented by the methods mentioned above into hidden features $\mathbf{h}_i^a=f_a(\mathbf{x}^a_i)$ and $\mathbf{h}_i^b=f_b(\mathbf{x}^b_i)$.
They are then projected into the embedding space using the projector $g$ and normalized as $\left(\mathbf{z}^a_i=g(\mathbf{h}_i^a) / ||g(\mathbf{h}_i^a)||, \mathbf{z}^b_i=g(\mathbf{h}_i^b) / ||g(\mathbf{h}_i^b)||\right)$. The projector is shared by all three modalities.
We have also tried using separate projectors but it yields worse performance.
The constrastive loss \cite{oord2018representation, chen2020simple} is used to push positives closer while negative ones are pushed further away in the embedding space. The loss function for $a\rightarrow b$ is defined as
%
\begin{equation}\label{eq1}
L^{a\rightarrow b}_{i} = -\textup{log}\frac{e^{( {\mathbf{z}^a_i}^T \mathbf{z}^b_i ) / \tau} }{\sum_{j=1}^N \left(\mathds{1}_{[j \ne i]} e^{\left(  {\mathbf{z}^a_i}^T \mathbf{z}^a_j  / \tau \right)} + e^{\left(  {\mathbf{z}^a_i}^T \mathbf{z}^b_j  / \tau \right) }\right) }
\end{equation}
where $\tau$ denotes the temperature parameter, $\mathds{1}_{[j \ne i]} \in \{0, 1\}$ is the indicator function evaluating to 1 iff $j \ne i$, and $N$ stands for the batch size.
Both intra- and inter-modality negatives are used, and there are $2N-2$ negative pairs in total.
The overall loss for the interaction between $a$ and $b$ is computed by summing $L^{a\rightarrow b}_i$ and its symmetric counterpart $L^{b\rightarrow a}_i$ across all positive pairs in the batch as $L^{ab}= \sum_{i=1}^N ( L_i^{a\rightarrow b}+L_i^{b \rightarrow a})$. The final loss is 
\begin{equation}\label{eq3}
  L = L^{vs} + L^{vw} + L^{sw}.
\end{equation}
Once it is trained, we use the output from the waveform or spectrogram encoder ($\mathbf{h}^w$ or $\mathbf{h}^s$) for downstream tasks. The video network is only used during training.












\section{Experimental Setting}

\subsection{Pretraining}
We pretrain our models on AudioSet \cite{gemmeke2017audio} sampled at 16 kHz. This dataset contains 2 million 10-second segments each labelled by multiple audio event labels from an ontology of 527 classes. The distribution of classes is highly imbalanced,
and training with a class-balanced dataloader is very effective in the supervised setting \cite{kong2020panns}. We do not use the labels during training, and hence we use regular sampling of the training data.

Unless specified otherwise, models are trained over 400k steps using a batch size of 4096 on a Cloud TPU v3 Pod slice with 32 cores.
We use the Adam optimizer~\cite{kingma2014adam}, first linearly increasing the learning rate from $0$ to $10^{-4}$ over 5k steps, and then decaying it following a cosine schedule \cite{loshchilov2016sgdr} down to $0$. We randomly crop 3-second windows for training. The video is sampled with a frame rate of 5 frames per second with resolution $50\times50$.
To be consistent with previous works \cite{saeed2020contrastive, wang2021multi},
for AudioSet downstream experiments, we pretrain and finetune our models using log-mel spectrograms with 80 features extracted by a window size of 20 ms and a stride of 10 ms; for other downstream tasks, we pretrain and finetune models using log-mel spectrograms with 64 features extracted by a window size of 25 ms and a stride of 10 ms. The maximum spectrogram shift size is $10$ and the example mixing ratio $\alpha$ is sampled on the fly from the $\beta(5, 2)$ distribution. All our models are trained with a temperature $\tau$ of 0.1, and the projector $g$ is a MLP that has 1 hidden layer of size 512 and ReLU nonlinearity.

\subsection{Downstream tasks}
We evaluate the representations on Audioset.
Following standard protocols \cite{jansen2018unsupervised, jansen2019coincidence, wang2020contrastive, alayrac2020self, wang2021multi}, we evaluate the performance of a shallow classifier trained in a supervised fashion on the frozen representations. Specifically, the classifier is a MLP with one hidden layer of size 512. Two batch normalization layers are used respectively after the representations and after the hidden layer. A ReLU non-linearity is applied after the second batch normalization. We train the classifier using Adam~\cite{kingma2014adam} with an initial learning rate of $2\times10^{-4}$ that decays following a cosine schedule over 30 epochs. Audio mixing and spectrogram shifting are used at training time. During evaluation, we equally split each sample into 10 sub-clips of 3 seconds, and average logits over sub-clips to obtain an overall score for the sample. We report the mean average precision (mAP) on the test set, together with the area under the curve (AUC) and d-prime as complementary metrics \cite{kong2020panns}.

We also evaluate the generality of our representations on a variety of downstream tasks previously studied in \cite{tagliasacchi2020pre, saeed2020contrastive}. For speaker identification, we consider 100 hours of the train-clean subset of Librispeech \cite{panayotov2015librispeech} with 251 speakers, and a larger one called VoxCeleb \cite{nagrani2017voxceleb} containing 1251 speakers. We employ the Speech Commands datasets (V1 \& V2) \cite{warden2018speech} for keyword spotting. Two tasks are taken from the DCASE2018 challenge: acoustic scenes classification \cite{heittola2018tut} and birdsong detection \cite{stowell2019automatic}. MUSAN \cite{snyder2015musan} is used for detection of music, speech, or noise. Moreover, we use VoxForge \cite{maclean2018voxforge} for language detection and Nsynth \cite{pmlr-v70-engel17a} for music instrument classification.
For all these tasks, we follow the linear evaluation protocol of \cite{saeed2020contrastive} and train the linear classifier on top of the frozen features with 1-second crops.
We do not apply further audio augmentations during training.
We use the Adam optimizer~\cite{kingma2014adam} with a learning rate of $2\times10^{-4}$.
At test time, we split the clip into non-overlapping 1-second sub-clips and average the scores across sub-clips.

\subsection{Models}

Three backbone networks are used under our framework.
For video, we use the TSM-ResNet50 (TSM-50) architecture \cite{lin2019tsm} as the backbone. For the waveform format, we use the Res1dNet31 \cite{kong2020panns}. Because it was originally introduced in the supervised setting, we remove its last two fully connected layers and directly learn the pooled outputs from the last Res1d block.
For the spectrograms, to be consistent with \cite{wang2021multi,kong2020panns}, we use CNNN14 \cite{kong2020panns} for the AudioSet experiments.
We employ EfficientNet-B0 \cite{tan2019efficientnet} for other downstream tasks to compare fairly with \cite{saeed2020contrastive, zeghidour2021leaf}, and adjust the feature dimension of TSM-ResNet50 and Res1dNet31 accordingly from 2048 to 1280. Similarly, the fully connected layers in both cases are removed.


\begin{table}[t]
    \centering
    \caption{\textbf{Modalities study}. We provide the test mAP of both the log-mel (CNN14) and waveform networks (Res1dNet31) on the downstream AudioSet task.}
    \begin{tabular}{ccc}
    \toprule
        Input modalities & Log-mel net & Waveform net\\
        \hline
        S, W & $36.1$ & $33.6$\\
        S, V & $38.5$ & -\\
        W, V & - & $35.0$\\
        S, W, V & $\mathbf{39.7}$ & $\mathbf{37.7}$\\
        \bottomrule
    \end{tabular}
    \label{tab:formats}
\end{table}

\begin{table}[t]
    \centering
    \caption{\textbf{Image resolution \& video input duration}. We show the mAP of the log-mel network (CNN14) on the downstream AudioSet task. The image size ablation is done with a smaller batch size of 512 to fit high-resolution models into the memory.}
    \begin{tabular}{ccccccccccc}
    \toprule
        Image size & 16 & 32 & 50 & 64 & 128 & 200 \\
        \midrule
        mAP & $35.0$ & $36.0$ & $\underline{36.3}$ & $36.3$ & $\mathbf{36.5}$ & $\mathbf{36.5}$    \\
        \bottomrule
    \end{tabular}
    \begin{tabular}{ccccccccc}
    \toprule
        Input duration (s) & 0.6 & 1.8 & 3 & 4.2 & 5.4 \\
        \midrule
        mAP & $33.0$ & $38.1$ & $\underline{39.7}$ & $\mathbf{39.8}$ & $39.0$   \\
        \bottomrule
    \end{tabular}
    \label{tab:image_size}
\end{table}


\begin{table}[t]
    \centering
    \caption{\textbf{Example mixing}. We show the mAP of the log-mel network (CNN14) trained with different batch sizes with or without mixing, and under different distributions for the mixing ratio.}
    \begin{tabular}{ccccccccc}
    \toprule
        Batch size & 512 & 1024 & 2048 & 4096 \\
        \midrule
        No example mixing & $38.2$ & $38.1$ & $37.4$ & $37.4$  \\
        + audio mixing    & $36.6$ & $38.4$ & $39.3$ & $39.5$  \\
        + video mixing    & $36.3$ & $38.0$ & $39.3$ & $\mathbf{39.7}$  \\
        \bottomrule
    \end{tabular}
    \begin{tabular}{cccc}
    \toprule
        Mixing ratio $\alpha$ & $\beta(5, 5)$ & $\beta(5, 2)$ & $\beta(5, 1)$ \\
        \midrule
        mAP & $35.5$ & $\mathbf{39.7}$ & $38.9$    \\
        \bottomrule
    \end{tabular}
    \label{tab:batch_size}
\end{table}

\begin{figure}[t]
  \centering
  \includegraphics[width=0.78\linewidth]{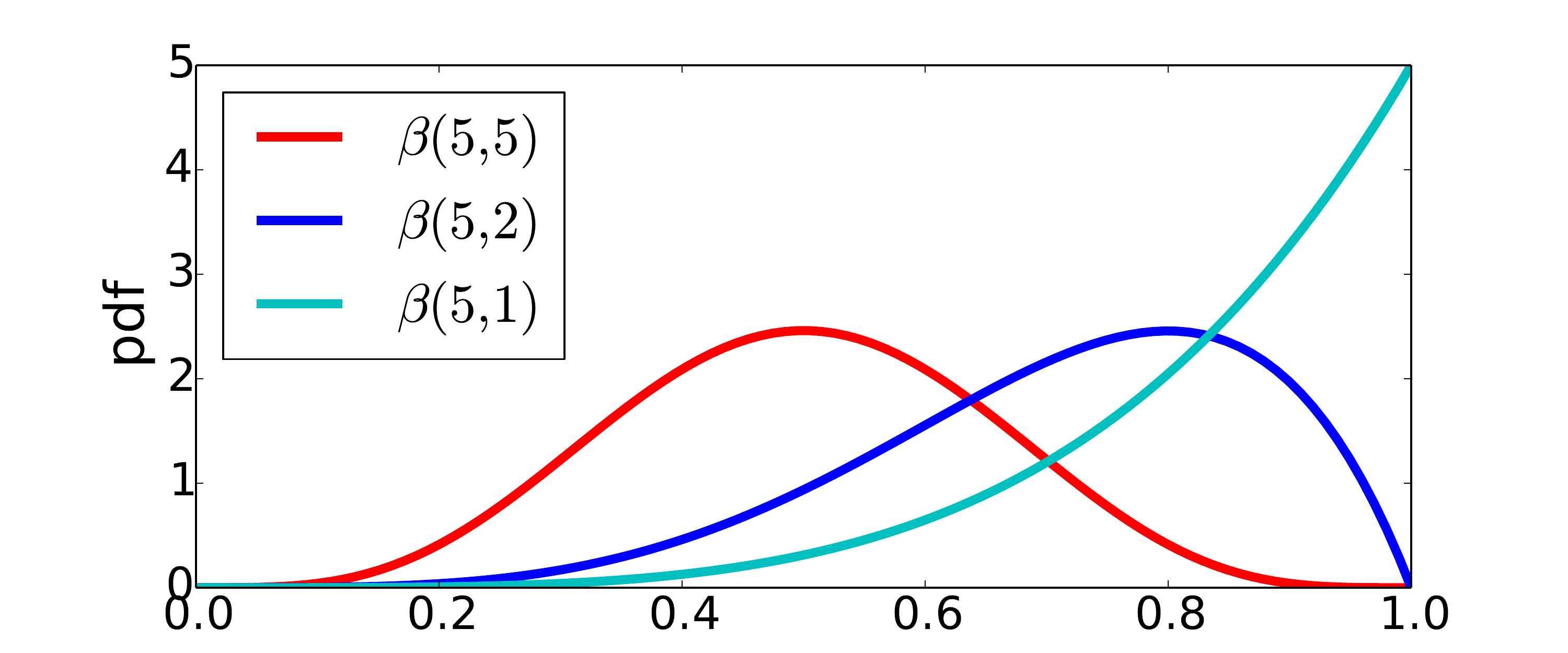}
  \caption{Density functions of the $\beta$ distributions we consider for the mixing ratio $\alpha$.}
  \label{fig:beta}
\end{figure}


\section{Results}

In this section, we give and analyze the results of our experiments. We start by conducting ablations to identify the important factors of our framework, and then compare our models to the unsupervised and supervised state-of-the-art on the AudioSet benchmark. Last but not least, we present the generalization performance of our models on diverse audio tasks.

\subsection{Ablation study}

\noindent \textbf{Benefits of using video:} In Table \ref{tab:formats}, we show the performance of both the log-mel (CNN14) and waveform networks (Res1dNet-31) trained with different training modalities and formats. Similar to \cite{wang2021multi}, we find that the log-mel networks outperform their waveform counterparts under all circumstances. More importantly, we observe that cross-modal training is advantageous over the unimodal multi-formats method presented in \cite{wang2021multi}. The mAP of the log-mel network improves from 36.1 to 38.5 by replacing the second waveform branch with video. Similar gain is found by replacing the log-mel branch with video. Combining all three branches as input, namely, video, log-mel, and waveform, further improves the mAP of the log-mel network to 39.7. It is clear that the information brought by video greatly improves the learned audio representations.
\\[0.2pt]

\noindent \textbf{Multimodal training with low image resolution and different timespan:} The video modality being higher-dimensional than audio, it is important to understand what is the impact of the spatio-temporal resolution of the video on learning audio features. We first look at image resolutions in Table~\ref{tab:image_size}, and find that the gain is minimal going beyond images of size $50\times50$, only increasing the mAP from 36.3 to 36.5 when the resolution increases to $200\times200$. This suggests that to learn good audio representations, it is sufficient to rely on high-level supervisory signals present in the video.

We then focus on the time dimension. We keep the number of image frames as 15 and change the video duration by varying the frame rate. It is shown in Table~\ref{tab:image_size} that the performance of the audio representations peaks around an input duration of $3$ to $4.2$ seconds. Therefore, we use 3-second long videos of resolution $50\times50$, considerably smaller than the spatio-temporal resolution commonly used for video tasks \cite{miech2020end, alayrac2020self} ($200\times200$ images and 32 frames).
\\[0.2pt]

\noindent \textbf{Example mixing and scaling up learning:} In Table~\ref{tab:batch_size}, we jointly investigate the impact of batch size and example mixing. If there is no mixing, the performance goes down when we increase the batch size.
If we add audio mixing, which simulates various forms of background noise, the models are trained to be invariant to the distractors, and presumably learn to focus more on the foreground content. Better representations are learned in this way and the performance nicely increases as a function of the batch size. Moreover, video mixing further improves the performance in the large-batch regime.

We also look at the impact of the example mixing ratio $\alpha$ in Equation~\ref{eq2}. We consider sampling this parameter from three kinds of $\beta$ distributions whose density functions are shown in Figure~\ref{fig:beta}, representing different levels of strength for the additive distractor. We find it best to sample the ratio from the $\beta(5, 2)$ distribution whose density function peaks around $0.8$.

\subsection{Comparing to the SOTA}

\begin{table*}[h]
  \caption{\textbf{Comparing to the state-of-the-art on AudioSet}. For unsupervised models, the evaluation is done by training a downstream shallow classifier on frozen representations.
  For fair comparison with supervised methods, in addition to mAP, we also report results when fine-tuning using a class-balanced dataloader (bal.) \cite{kong2020panns}.
  All models are (pre)trained on AudioSet.
  }
  \label{tab:benckmark}
  \centering
  \begin{tabular}{ l c c c c c c  c}
    \toprule
    \textbf{Model} & \textbf{Training modalities} & \textbf{Downstream net}   & \textbf{Audio format} &  \textbf{mAP} & \textbf{mAP (bal.)} & \textbf{AUC (bal.)} & \textbf{$\mathbf{d^{\prime}}$ (bal.)} \\
    \midrule
    \emph{Supervised} & & \\
    PANNs \cite{kong2020panns} & W & Res1dNet31 & W  & - & $36.5$ &  $0.958$ & $2.44$~~~\\
    PANNs  \cite{kong2020panns} & S & CNN14 & S & - & $\mathbf{43.1}$ & \underline{$0.973$} & \underline{$2.73$}~~~\\
    LEAF \cite{zeghidour2021leaf}  & S & CNN14 & S & - & - & $\mathbf{0.974}$ & $\mathbf{2.74}$~~~\\
    \midrule
    \emph{Unsupervised} & & & \\
    Triplet \cite{jansen2018unsupervised} & S & ResNet-50 & S  & $24.4$ &- &- &-~~~\\
    $L^3$ \cite{arandjelovic2017look} & V, S  & ResNet-50 &  S    & $24.9$ &- &- &-~~~\\
    CPC \cite{wang2020contrastive} & W  & TDNN & W & $27.7$ &- &- &-~~~\\
    $C^3$ \cite{jansen2019coincidence} & V, S & ResNet-50&  S   & $28.5$ &- &- &-~~~\\
    MMV \cite{alayrac2020self} & V, S & ResNet-50 & S        & $29.7$ &- &- &-~~~\\
    MF \cite{wang2020contrastive} & S, W & Res1dNet31 &  W & $35.5$ & - &- &- ~~~\\
    MF \cite{wang2020contrastive} & S, W & CNN14 & S & $36.8$ & - &-&-~~~\\
    \midrule
    Ours & V, S, W & Res1dNet31 &  W    & \underline{$37.7$} & $40.5$ & $0.972$ & $2.70$~~~\\
    Ours & V, S, W & CNN14 &  S        & $\mathbf{39.7}$ & \underline{$42.4$} & \underline{$0.973$} & \underline{$2.73$}~~~\\
    \bottomrule
  \end{tabular}
  
\end{table*}

\omitme{In Table~\ref{tab:benckmark}, we compare our models to the state-of-the-art on AudioSet. Our audio models outperform all previous unsupervised methods. It performs significantly better than MMV \cite{alayrac2020self}, the previous best audiovisual model. Counterintuitively, our waveform network, commonly known to be inferior to log-mel models, performs with a mAP of $37.7$, which is better than the previous SOTA log-mel network at $36.8$.}

In Table~\ref{tab:benckmark}, we compare our models to the state-of-the-art on AudioSet. Our spectrogram model (CNN14) outperforms all previous unsupervised methods with an mAP of $39.7$. Despite that waveform networks are known to be inferior to spectrogram models, our Res1dNet31, performing at $37.7$, is even better than the previous best spectrogram network.

\omitme{
Furthermore, our models are even comparable to supervised models.
We train the downstream MLP classifier with an increased hidden layer size of 2048. The total number of parameters in this shallow classifier and the frozen pretrained model is equal to the supervised ones reported in \cite{kong2020panns, zeghidour2021leaf}.
A class-balanced dataloader is also used in our case \cite{kong2020panns}. Our model achieves a mAP of 42.4 using \emph{frozen} features. This closes the gap between unsupervised and supervised methods. Meanwhile, there is almost no difference on AUC and d-prime. However, comparing to 43.1 mAP using supervised learning, the convolutional blocks in our model are not trained with the balanced classifier. In addition, our waveform model outperforms the supervised SOTA. These findings show that the video modality provide strong signals to train audio networks even in the absence of labels.
}

Our models are even comparable to state-of-the-art supervised models \cite{kong2020panns, zeghidour2021leaf}.
To have a fair comparison to these models, we train the downstream MLP classifier with one hidden layer of size 2048: the total number of parameters in this classifier and the frozen pretrained model is the same as that of the supervised model.
A class-balanced dataloader is also used, but in our case we only use it to train the downstream classifier.
As a result, our CNN14 performs similarly to same model trained in a supervised fashion. Comparing to PANNs \cite{kong2020panns}, there is no difference on AUC and d-prime: both are at $0.973$ and $2.73$, respectively. The mAP is only slightly worse (42.4 vs 43.1). LEAF \cite{zeghidour2021leaf} uses a more expressive learnable frontend instead of spectrograms but it is just marginally better. Meanwhile, our Res1dNet31 even outperforms its supervised counterpart, producing a new state-of-the-art for waveform models with an mAP of $40.5$. These findings show that the video modality provide strong signals to train audio networks even without labels.

\subsection{Generalization to other audio downstream tasks}


\begin{table}[t]
    \centering
    \caption{\textbf{Generalization to other tasks}. We show test accuracy (\%) on different downstream tasks trained with a linear classifier on top of the frozen features outputed from our pre-trained network, comparing to supervsied and unsupervsied baselines. All methods are based on EfficientNet-B0.}
    \begin{tabular}{lccc}
    \toprule
        Task & COLA \cite{saeed2020contrastive} & Ours & \spv{Sup. \cite{zeghidour2021leaf}}  \\
        \hline
        Speaker Id. (Librispeech) & $\mathbf{100.0}$ & $99.6$ & - \\
        Speech commands (V1) & $71.7$ & $\mathbf{80.5}$ & \spv{$93.4$} \\
        Speech commands (V2) & $62.4$ & $\mathbf{82.2}$ & - \\
        Acoustic scenes & $\mathbf{94.0}$ & $90.4$ & \spv{$99.1$} \\
        Speaker Id. (VoxCeleb) & $29.9$ & $\mathbf{38.2}$  & \spv{$33.1$} \\
        Birdsong detection & $77.0$ & $\mathbf{80.0}$ & \spv{$81.4$} \\
        Music, speech \& noise & $99.1$ & $\mathbf{99.6}$ & - \\
        Language Id. & $71.3$ & $\mathbf{79.0}$  & \spv{$86.0$} \\
        Music instrument & $63.4$ & $\mathbf{68.3}$ & \spv{$72.0$} \\
        \hline
        Average & $74.3$ & $\mathbf{79.8}$  &  - \\
        \bottomrule
    \end{tabular}
    \label{tab:generalization}
\end{table}

In Table~\ref{tab:generalization} we study how our model generalizes to other audio tasks. Our model outperforms COLA \cite{saeed2020contrastive} on 7 out of 9 tasks with the same EfficientNet-B0 architecture and downstream evaluation setting. It has a mean accuracy of $79.8\%$. The results suggest video is a reliable additional supervisory signal for a wide range of audio tasks.

We observe that speech-related tasks benefit more from the additional visual information brought by our framework. Notably, there is significant improvement on the speaker identification task on VoxCeleb from an accuracy of $29.9\%$ to $38.2\%$, which even outperforms the supervised benchmark of $33.1\%$ from \cite{zeghidour2021leaf}. Similar improvement is also observed on language identification. Meanwhile, there is a significant gain on the keyword spotting tasks on Speech Commands v1 and v2, increasing from $71.7\%$ and $62.4\%$ to $80.5\%$ and $82.2\%$, respectively. We also observe improvements on birdsong detection and musc, speech, and noise. There is a slight decrease on Librispeech speaker identification and acoustic scenes classification. However, on average our model is $5.5\%$ better than COLA without additional tricks such as the bilinear similarity measure.

\section{Conclusions}

In this paper, we investigate learning general audio representations with video, spectrograms, and raw waveforms.
We find that good audio features do not require signals from high-resolution images. Meanwhile, we observe that example mixing makes the pretext task considerably harder. Both prove to be beneficial for improving the quality of the learned representations.
As a result, our models set a new state-of-the-art on AudioSet, as well as a broad class of downstream tasks.
Our work shows audio models learned in an unsupervised fashion are comparable to their supervised counterparts as the video provides strong signals for learning, paving the way for learning audio representations with larger unlabeled video datasets.

\section{Acknowledgements}

The authors would like to thank Marco Tagliasacchi and Neil Zeghidour for their help with the downstream tasks.

\bibliographystyle{IEEEtran}

\bibliography{mybib}

\end{document}